\documentclass[%
 reprint,
 amsmath,amssymb,
 aps,
]{revtex4-1}

\usepackage{graphicx}
\usepackage{dcolumn}
\usepackage{bm}

\begin{document}

\preprint{APS/123-QED}

\title{A relaxation constant in the folding of thin viscoelastic sheets}

\author{Kasra Farain}

\affiliation{Department of Physics, Sharif University of Technology, P.O. Box 11155–9161, Tehran, Iran\\
 farain@physics.sharif.edu}

\begin{abstract}
If one folds a thin viscoelastic sheet under an applied force, a line of plastic deformation is formed which shapes the sheet into an angle. We determine the parameters that define this angle experimentally and show that, no matter how much load one applies, it is impossible to make angles less than a certain minimum angle in a definite time. Moreover, it is shown that regardless of whether the sheet is released freely afterward or kept under the load, a logarithmic relaxation process follows the first deformation. The slope of this logarithm is the same in both conditions and depends neither on the applied force nor on the thickness of the sheet, which indicates it is directly a probe of the molecular mobility of the material. This intrinsic relaxation constant was measured 0.01 and 5.7 for Mylar and paper sheets, respectively. It is also suggested that the observed minimum angle of folding can be defined as a characteristic index for the plasticity of different materials.

\end{abstract}

\pacs{Valid PACS appear here}
\maketitle



Folding is at first sight a seemingly trivial way of making multilayer stacks or three-dimensional objects from thin sheets. However, folding problems arise in a variety of living organisms such as biological membranes, insect wings, and the cortical brain structure; as well as in artistic and technological applications ranging from decorative art and fashion to space solar panels. For instances, the shape of viral shells are determined by the energies of folding \cite{1, 2}; the cerebral cortex expansion occurs with increasing degrees of folding of the cortical surface \cite{3}; creasing properties of fabrics like crease-resistance and recovery are very important for designing suits, skirts and ladies' dresses; and engineers seek to design series of solar panels which can change between folded stowed and planar configurations. Other examples which more clearly require a better understanding of folding and have recently attracted a great deal of attention are origami designed structures \cite{5, 6, 7, 8} and crumpled sheets \cite{10, 11, 12}, consisting respectively of ordered and random networks of creases created in a sheet. Mechanical properties of such systems are determined by not only the network but also by the response of each of the building block creases which can be considered as an elastic hinge of specific stiffness connecting flexible panels \cite{13}. Furthermore, next-generation soft robots \cite{14, 15} and wearable electronics \cite{16} include thin 2D elastomeric parts that undergo continual bending and folding during use. Understanding the structural and geometrical changes, force production, as well as non-linear and time-dependent responses of these folding parts are enormously important for precise and efficient control of them, especially when small forces are exploited \cite{17}.

In this letter, we experimentally study folding of thin viscoelastic sheets and investigate how the remained plastically-deformed crease behave afterward. Imagine a piece of paper bent gently and put under some slowly increasing force similar to the inset of Fig 1. The elastic energy of bending, first spread smoothly throughout the curved region, suddenly concentrates in a strongly bent edge \cite{18}. In other words, at some point the maximum stress passes the yield point of the material and triggers a non-continuous self-accelerating process (geometry of the applied force) which makes permanent changes in the structure. A viscoelastic polymer, when subjected to an applied stress, shows a combination of time-dependent irreversible viscous flow and immediate recoverable elastic strain \cite{19, 20}. When a crease is created in a viscoelastic sheet after pressing by a certain load for a short time interval, the irreversible viscous flow continues under the effect of the elastic response even after releasing. This is, in a sense, a perturbation in a disordered molecular system which relaxes towards an equilibrium (or quasi-equilibrium) state over several decades of time \cite{21}. Similar slow relaxations have been observed in a wide variety of out-of-equilibrium disordered systems such as glassy polymers below the glass transition temperature \cite{22, 23}, amorphous metals \cite{24, 25, 26}, granular media \cite{27, 28}, frictional interfaces \cite{29}, crumpled sheets \cite{11, 30}, and electron glasses \cite{31}. Apart from theoretical implications, due to their vast use in technology, slow time-dependent structural or geometrical changes are particularly important in the case of polymer glasses. Here, in addition to presenting a thorough phenomenological description of folding using Mylar sheets, we introduce a simple and fast way for perturbing a polymer glass to an out-of-equilibrium state, which follows by a robust logarithmic relaxation. The time origin of this logarithm corresponds to the perturbation time. Moreover, it is shown that, independent of the folding parameters and external geometrical constraints on the fold afterward, the logarithmic process evolves with a constant rate which is a characteristic of the material. The slopes of similar logarithmic changes are widely used to compare recovery dynamics of polymers \cite{32}.

\begin{figure}
	\begin{center}
		\includegraphics[scale=0.29]{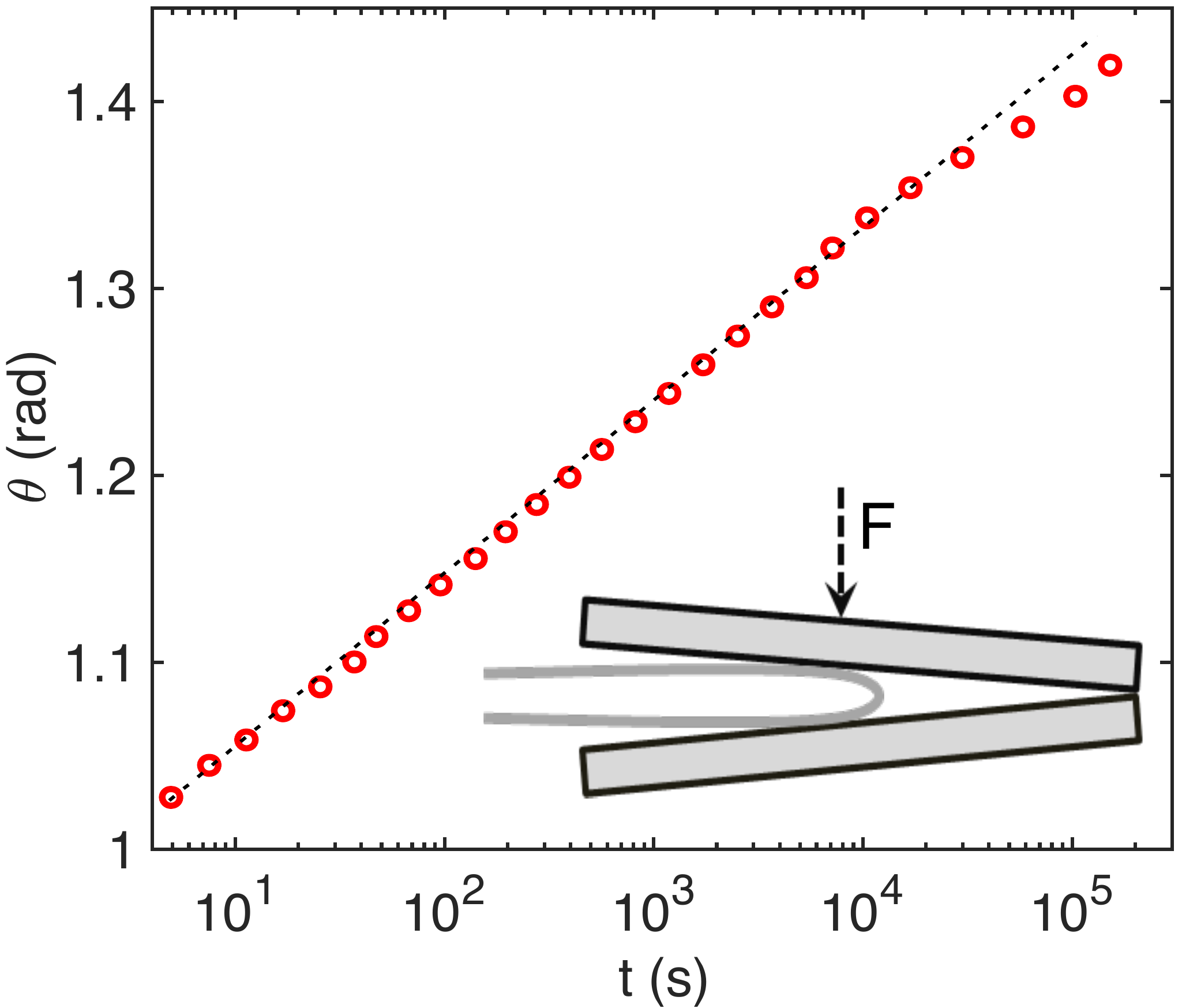}
	\end{center}
	\vspace{-.5cm}
	\caption{Time evolution of the angle of a typical crease in the 0.25 mm-thick Mylar sheet. The crease has been produced at $t=0$. The inset shows a schematic of the fold preparation.}
	\label{f:numeric}\end{figure}

Commercial Mylar sheets of different thicknesses of 0.08, 0.10, 0.17 and 0.25 mm were cut into 15 mm $\times$ 25 mm rectangles. These pieces were then bent smoothly by hand along the short edge (the crease length of 15 mm) and pressed under defined forces between two flat metallic plates for 3 seconds. The plates were hinged at one side, which guarantees that all the force is concentrated on the crease line when they are closed. Moreover, to avoid any extra shock to the folded edge, the plates were closed by a handle slowly. After releasing, the folded pieces settle in an initial angle instantly and then, as also observed by others \cite{21}, start opening in a perfect logarithmic way over several decades in time (Fig. 1). One side of the samples was held by a gripper and the other side was freely opening parallel to the ground. The samples were illuminated from the side and observed with a camera. The logarithmic unfolding can be stated by the following equation
\begin{equation}
 \theta = a\log(t-t_{0})+b,
\end{equation}
where $a$ and $b$ are constants. $t_{0}$ corresponds to the instant that the crease has been created. Knowing the logarithmic dependence of the angle on $t - t_{0}$, $t_{0}$ can be also extracted from the unfolding data within a few seconds accuracy. This means that by simply observing the relaxation of a folded sheet, one can say when the folding has happened. However, it should be pointed out that the logarithmic relationship is not valid in the immediate vicinity of $t_{0}$, since it diverges. Additionally, more waiting in the loading stage will lead in a larger deviation from the logarithmic behavior in the starting seconds, as previously reported by Thiria and Adda-Bedia for the relaxation of the force produced by a creased sheet \cite{21}. The small deviation in the linear behavior after $2\times 10^4$ s in Fig. (1) probably happens when the generated force in the creased region is comparable with other mechanisms such as vibrations in the system. This deviation starts after $\sim 10^3$ s when the direction of the sample is so that the gravitational force is also working on the system. The discussion presented here may also help to understand the relaxation behavior of crumpled sheets, which consist of a random pattern of creases with different ages and strengths through consecutive crumpling at different times \cite{11}.

\begin{figure}[hbt]
	\begin{center}
		\includegraphics[scale=0.29]{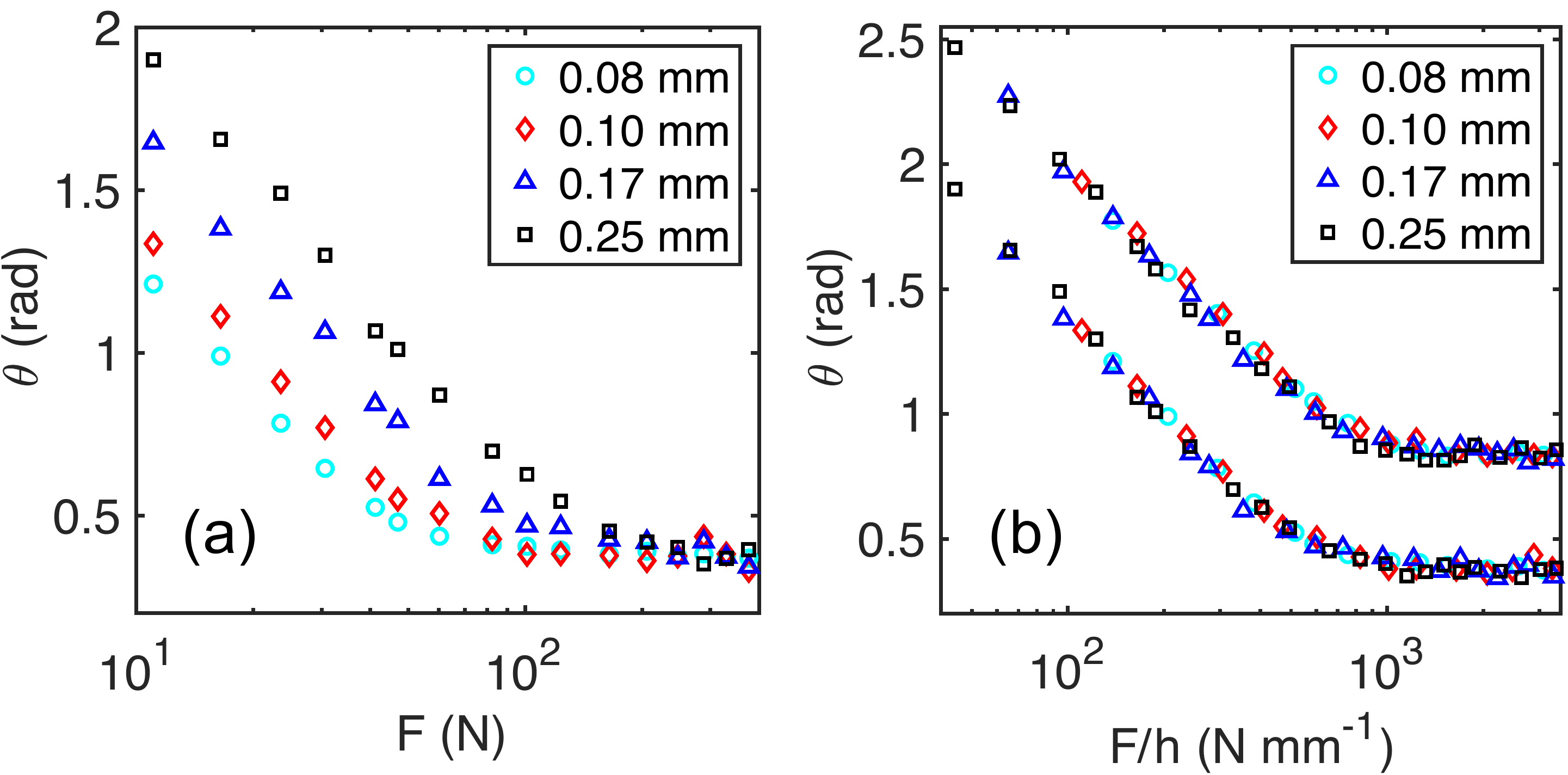}
	\end{center}
	\vspace{-.5cm}
	\caption{(a) The crease angle of Mylar sheets with four different thicknesses, $t_{i}=3$ s after folding, as a function of the applied force. All the curves finally converge into the same limit angle. (b) The angle of the folded sheets at $t_{i}=3$ s (lower curve) and $t_{f}=48$ hours (upper Curve) plotted as a function of the applied force over the thickness of the sheets. The data points for the different thicknesses collapse onto each other. The vertical distance between $t_{i}$ and $t_{f}$ curves remains constant throughout the logarithmic and saturation (after  $10^3$ N mm$^{-1}$) parts.}
	\label{f:numeric}\end{figure}

When one folds a sheet of paper in two, they usually press the fold edge strongly to minimize the returning of the second layer. But, to what extent does this extra pressing help? Figure 2(a) shows the crease angle in the folded Mylar pieces of different thicknesses as a function of the load acted on them, $t_{i}=3$ s after folding. At first, the obtained angle decreases logarithmically with the applied load. Then, it reaches a limit angle and further increasing of the applied load does not affect the crease anymore. Additionally, all the four curves for the different thicknesses of Mylar finally converge into the same limit angle. Therefore, it is not possible, by applying larger loads, to make angles less than a certain minimum angle in the Mylar sheets. 

From the elasticity theory for small deflections of thin plates, we know the bending rigidity as $B={Eh^3}/{12(1-\nu^2)}$ where $h$, $E$ and $\nu$ are the thickness of the plate, the modulus of elasticity and the Poisson ratio, respectively \cite{33}. Considering the bending rigidity as the moment per unit length of the crease per unit of curvature ${Fh}/{L(1/h)}$ and compare it with the above expression, one can expect $F/LhE$ to be the relevant dimensionless quantity for describing the angle caused by folding. The lower curve in Fig. 2(b) shows the data of Fig. 2(a) when the applied load is scaled by the thicknesses of the sheets. This plot confirms that the folding force appears correctly as $F/h$ in the governing relation. The upper curve in Fig. 2(b) presents the angle of the same samples after $t_{f}=48$ hours. Knowing $\theta$ at two different times is enough to find the constant $a$ and $b$ of Eq. (1):
\begin{equation}
\theta = \theta_{i}+\dfrac{\theta_{f}-\theta_{i}}{\log(\dfrac{t_{f}}{t_{i}})}\log(\dfrac{t}{t_{i}}),
\end{equation}
where we have supposed $t_{0}=0$. Interestingly, as observed in Fig. 2(b), the lower and upper curves have the same slope and $\theta_{f}-\theta_{i}$ is a constant throughout the logarithmic region and, with a small change, in the saturation region (after $10^3$ N mm$^{-1}$). Therefore, the quantity $a={(\theta_{f}-\theta_{i})}/{\log({t_{f}}/{t_{i}})}$ depends neither on the applied force nor on the thickness of the sheets. Rearranging Eq. (2) as
\begin{equation}
\dfrac{\theta-\theta_{i}}{\log(\dfrac{t}{t_{i}})} = \dfrac{\theta_{f}-\theta_{i}}{\log(\dfrac{t_{f}}{t_{i}})},
\end{equation}
 one can see it does not depend on the specific choice of $t_{i}$ and $t_{f}$ either  and can be regarded as a material property. For the Mylar sheets used in this work, in the logarithmic area $a=0.124 \pm 0.005$, and in the limiting minimum angle or the maximum deformation part $a=0.098 \pm 0.005$ were obtained. Put concisely, although the crease angle, up until a minimum angle, scales with $\log(F/h)$ (which is represented by the value of $\theta_{i}$), but the relaxation toward equilibrium is independent of the applied pressing force and the thickness of the sheet, and happens with a constant rate (which is given by the slope $a$).

We also examined how the folding force needed to make an angle in a sheet is related to the length of the fold. Figure 3(a) shows the angle of 0.17 mm-thick Mylar pieces with three different widths (fold length) of 8.5, 15 and 20 mm at $t_{i}=3$ s versus the load acted on them. In Fig. 3(b), the data collapse onto a single curve when are plotted as a function of the force per unit length of the fold. Therefore, the folding force for creating a certain angle in a sheet is also proportional to the length of the fold and the rescaled force ${F}/{Lh}$  must be used to characterize folding. This is opposed to the current belief that generally supposes the folding energy is independent of the length of the fold \cite{10, 34}. It should be noted that the linear proportionality between the applied load and crease length is only captured before the saturation of the deformation.

\begin{figure}[hbt]
	\begin{center}
		\includegraphics[scale=0.29]{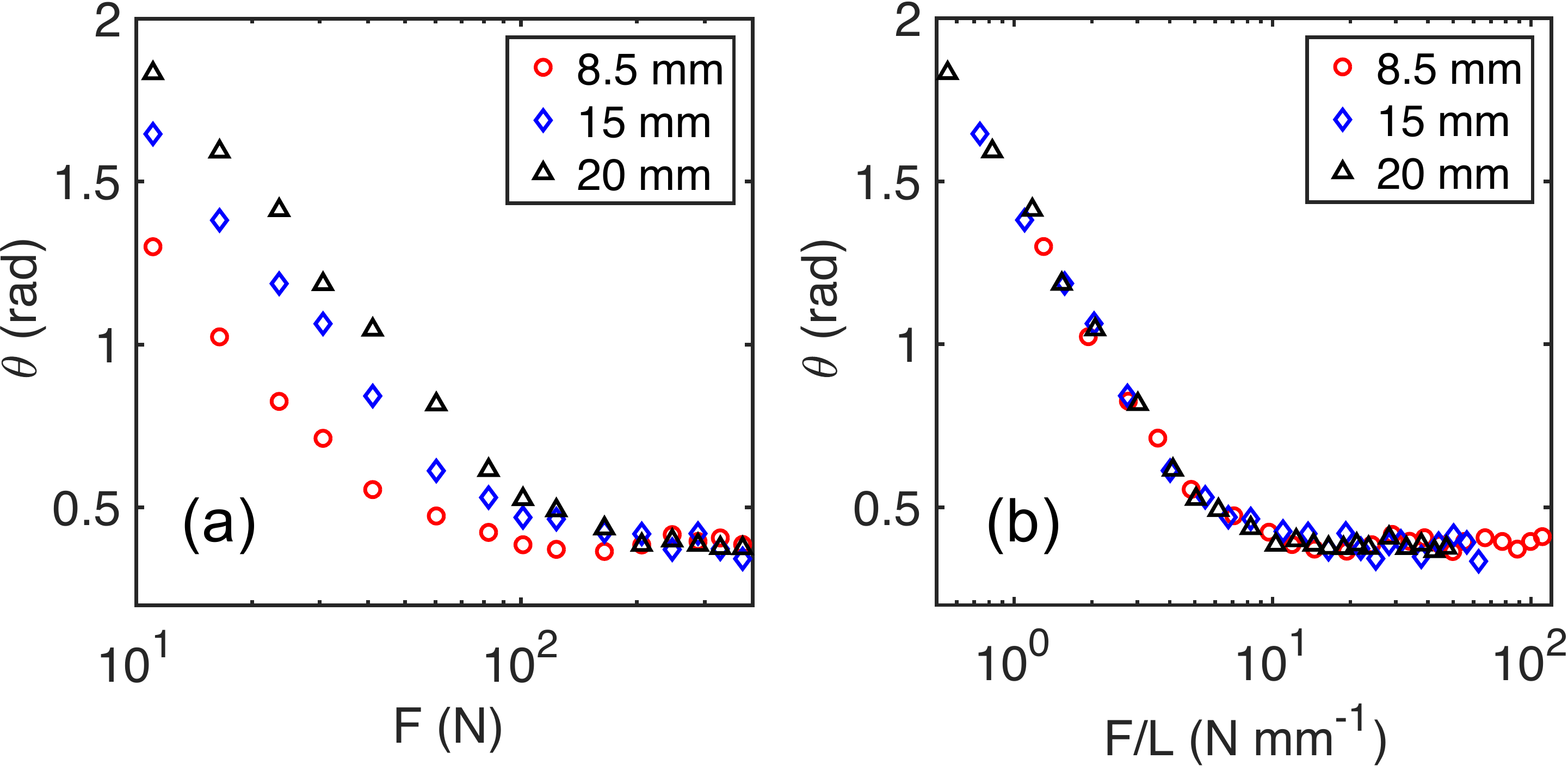}
	\end{center}
	\vspace{-.5cm}
	\caption{(a) The crease angle of 0.17 mm-thick Mylar pieces at $t_{i}=3$ s as a function of the applied pressing force for three different lengths of the crease, 8.5, 15 and 20 mm. (b) The data collapse onto each other when plotted as a function of the applied force per unit length of the crease.}
	\label{f:numeric}\end{figure}

One remarkable feature of the results presented in Fig. 2 and Fig. 3 is that there is a minimum angle which, no matter how much load one uses, it is impossible to fold a Mylar sheet more than that angle. This minimum angle, which is also independent of the thickness of the sheets and therefore is a characteristic of the material, is 50 degrees for Mylar, and for instance $\sim$ 36, 118, 0 degrees in the case of printing paper, polyethylene sheets and aluminum foils, respectively. These results suggest that the minimum fold angle may be utilized to define an index for plastic behavior of different materials.

\begin{figure}[hbt]
	\begin{center}
		\includegraphics[scale=0.29]{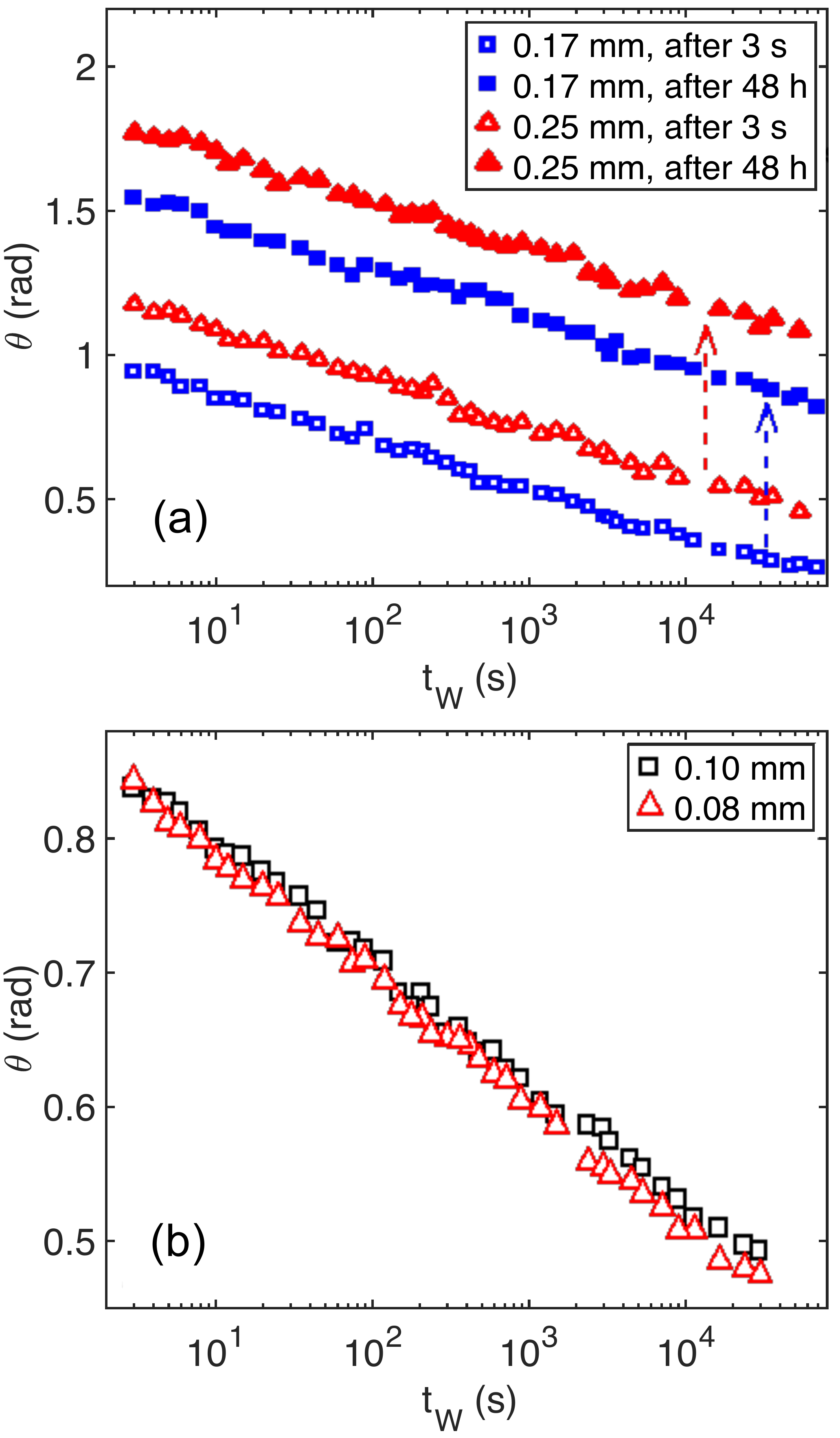}
	\end{center}
	\vspace{-.5cm}
	\caption{(a) The crease angle at  $t_{i}=3$ s and $t_{f}=48$   hours after removing the load, for Mylar sheets with thicknesses of 0.25 and 0.17 mm, length of 15 mm and the applied force of 40 N (which falls before the saturation area for both thicknesses), as a function of the loading time $(t_{W})$. (b) Same as part (a), but for Mylar sheets with thicknesses of 0.10 and 0.08 mm, the angle after $t_{f}=48$   hours, and the applied force of 125 N (which falls in the saturation part).}
	\label{f:numeric}\end{figure}

Furthermore, the effect of the waiting time under the load, $t_{w}$, on the final angle of the folded sheets was investigated. In Fig. 4(a), the measured angles at $t_{i}=3$ s and $t_{f}=48$ hours for Mylar sheets with thicknesses of 0.25 and 0.17 mm and width of 15 mm are plotted versus the different loading times. The applied force was 40 N, which for both thicknesses falls in the logarithmic part of the curves in Fig. 2 and Fig. 3.  As observed, the crease angle decreases also logarithmically with the loading time. The slope of this logarithmic decrease is the same for the angles at  $t_{i}=3$ s and $t_{f}=48$  hours for both thicknesses, and equals $0.16 \pm 0.02$. Figure 4(b) shows the same results for Mylar sheets with thicknesses of 0.10 and 0.08 mm when the loading force is high enough to fall in the saturated part (125 N). In this regime, the slope of the logarithmic decrease of the crease angle is $0.09 \pm 0.01$, which is the same as the slope of the logarithmic opening of a folded piece after releasing it $(0.098 \pm 0.005)$. The bigger slope observed in Fig. 4(a) might be justified with the macroscopically dynamic state (vibrations) of the bent Mylar pieces and smaller Hook’s constant of the configuration in that experiment.

After folding, the strongly bent region has been stressed beyond the yield point and undergone defects and sudden changes in the molecular arrangement. The new structure is out of equilibrium and immediately starts a relaxation process towards the maximum entropy state. Moreover, this relaxation takes place toward a molecular order which is dictated by the external geometrical constraints on the sample. In Fig. 4, when the folded Mylar pieces are kept under the load, the molecular activity evolves toward a state compatible with that pressed condition. After unloading, the samples continue relaxation toward an equilibrium consistent with the free state. Comparing the slope of the logarithmic unfolding and the crease angle changes with the loading time suggests that the relaxation mechanism is the same in both cases and does not depend on the external mechanical constraints on the system. To examine the general character of this result, we conducted the same experiments with printing paper and obtained $a=5.7 \pm 1.5$ for the relaxation constant in both folding and unfolding tests. 

In conclusion, we studied the relaxation mechanism in the folding and unfolding of thin viscoelastic sheets. In addition to describing how the crease angle changes with the thickness of the sheet, the crease length and the magnitude as well as the acting time of the applied pressing force, we showed that after the first instantaneous plastic deformation a slow relaxation process proceeds logarithmically. The rate of this process is the same when the sheet is freely opening or remains under the pressing force longer and is independent of the sheet thickness and the applied force. Therefore, it is identified as a material property which can be utilized as an experimental measure for assessing molecular mobility and stress relaxation rate in polymer glasses. Moreover, in a given time of loading the minimum achievable fold angle is limited and constant for all thicknesses of the same material, and can be a characteristic index for defining the plasticity of different materials.

\vspace*{0.3 cm}
The author would like to express his gratitude to Professor Shmuel Rubinstein of Harvard University for his support and stimulating discussions throughout this study.

\end{document}